# Experimental evidence of auxetic features in seismic metamaterials: Ellipticity of seismic Rayleigh waves for subsurface architectured ground with holes

Stéphane Brûlé[1], Stefan Enoch[2] and Sébastien Guenneau[2]

[1]Ménard, Chaponost, France

1, Route du Dôme, 69630 Chaponost

[2] Aix Marseille Univ, CNRS, Centrale Marseille, Institut Fresnel, Marseille, France

52 Avenue Escadrille Normandie Niemen, 13013 Marseille

e-mail address: stephane.brule@menard-mail.com

Structured soils with regular meshes of metric size holes implemented in first ten meters of the ground have been theoretically and experimentally tested under seismic disturbance this last decade. Structured soils with rigid inclusions embedded in a substratum have been also recently developed. The influence of these inclusions in the ground can be characterized in different ways: redistribution of energy within the network with focusing effects for seismic metamaterials, wave reflection, frequency filtering, reduction of the amplitude of seismic signal energy, etc. Here we first provide some time-domain analysis of the flat lens effect in conjunction with some form of external cloaking of Rayleigh waves and then we experimentally show the effect of a finite mesh of cylindrical holes on the ellipticity of the surface Rayleigh waves at the level of the Earth's surface. Orbital diagrams in time domain are drawn for the surface particle's velocity in vertical (x, z) and horizontal (x, y) planes. These results enable us to observe that the mesh of holes locally creates a tilt of the axes of the ellipse and changes the direction of particle movement. Interestingly, changes of Rayleigh waves ellipticity can be interpreted as changes of an effective Poisson ratio. However, the proximity of the source is also important to explain the shape of the ellipses. We analyze these observations in terms of wave mode conversions inside the mesh and we propose to broaden the discussion on the complexity of seismic wave phenomena in structured soils such as soils foundations and on the coupling effects specific to the soil-structure interaction.

## I. Introduction

Following two full-scale experiments on control of surface waves led in France in 2012 [1, 2], structured soils made of cylindrical voids or soft/rigid inclusions [3-7], have been coined seismic metamaterials [8]. One of the full-scale experiments was performed near Lyon, with metric cylindrical holes that allowed the identification of various effects such as a Bragg effect and the distribution of energy inside a grid, which can be interpreted as the consequence of an effective negative refraction index [8]. The pattern of the grid of holes was as follows: 20 m in width, 40 m in length) made of 23 holes (2 m in diameter, 5 m in depth, triangular grid spacing 7.07 x 7.07 m). This flat lens effect for surface seismic waves in [1] is reminiscent of what Veselago [9] and Pendry [10] envisioned for light in Electromagnetism. Bearing in mind the different ways of characterizing the effects of a structured ground in civil engineering, we decided to go a step further in the analysis of physical phenomena reported in [1] by showing their complexity. This opens new avenues for discussion and interpretation. We consider that we are at the beginning of a period that will demonstrate the significant role of buried structures (local geology, high density of foundations, etc.) on the free surface dynamic response. To illustrate the interaction of seismic waves with structured soils, researchers have to bring specific theoretical and original experimental approaches in particular because of the complexity of the wave propagation in the real Earth's surface layers ([1] and [2]). Except for a few cases of study with full numerical modeling, a complete study of both the structure and deep foundations (piles, shear-walls) is rather complex and usually the problem is split into two canonical sub-problems: kinematic and inertial interactions [11]. Kinematic interaction results from the presence of stiff foundation elements on or in soil, which causes motions at the foundation to depart from free-field motions. Inertial interaction refers to displacements and rotations at the foundation level of a structure that result from inertia-driven forces such as base shear and moment. New developments during this last decade are related to the interaction of structured soil with the seismic signal propagating in superficial Earth's layers [12] and then, linked to an active action on the kinematic effect described above.

In this article we propose new arguments, based on experimental facts, on the polarization change of Rayleigh surface waves. We show that the holes behave like resonators that can lead to a change of mode propagation inside the grid. To do this, we propose to change the conventional approach based on the distribution of seismic energy in the grid and to focus on the advantages brought by the simple polarization change of the surface waves on the design of buildings. Indeed, buildings are especially sensitive to horizontal displacement of the soil at the free surface [11].

## II. Ellipticity of Rayleigh waves

Surface waves are produced by body waves in media with a free surface and propagate parallel to the surface. For an epicenter remote from the zone of interest, the body waves emitted by the seismic source (shear and compression waves) are assimilated to plane waves because of their large radius of curvature. By essence, the amplitude of surface wave decreases quickly in depth. For our purpose, it is enough to consider an elastic, homogeneous half-space made of soils (Supplemental Figure 1). Equations governing the displacement of soil's particle are described in §1 of the Supplemental Material. Rayleigh waves are theoretically polarized in the vertical plane $(x, z)$ and we obtain an ellipse for the particle's motion with a vertical major axis and retrograde motion, opposite to that of wave propagation (Supplemental Figure 2). There is a value of $z = -0.19 \Lambda$ ($\Lambda = 2\pi/k$ is the wavelength inversely proportional to $k$ the wavenumber) for which $u_x$ is null, whereas $u_z$ is never null. At this depth, the amplitude of $u_x$ changes sign. For greater depths, the particle motion is prograde. With increasing depth, the amplitudes of $u_x$ and $u_z$ decrease exponentially, with $u_z$ always larger than $u_x$ (Supplemental Figure 3). We led a parametric study on the Poisson's ratio for the ellipse motion at free surface ($z = 0$). The size of the ellipse grows with the Poisson's ratio and we notice that the dimensions of the ellipses increase very quickly when ν tends towards 0.5 (FIG. 1). The geometric transformation is not just a homothety. In fact if one would like that the Poisson's ratio ν takes values between 0.01 and 0.499, the ratio $u_{z\,max}/u_{x\,max}$ should vary from 0.54 to 0.77. It is worthwhile noticing that controlling the Poisson's ratio opens an interesting avenue to cloaking, in a way analogous to what has been proposed with mechanical metamaterials at a micrometer scale [13]. Let us stress again that the above discussion is based upon a simplified soil model and most real Earth-motion records are more complex, due to the variation of ideal elastic half-space theory and the real soils (fine horizontal strata, increase of the density with depth, material damping describing all anelastic effects, etc.).

## III. Vertical point force

Lamb (1904) [14] extended Rayleigh's results, which were limited to the free propagation of waves, by finding the complete response of a homogeneous half-space to both a line and a point force [15, 16]. The rigorous solution of the ground surface displacement is expressed as a sum of the branch integrals and the residue of Rayleigh poles. The former corresponds to the contribution of body waves and the latter to that of Rayleigh waves. The evaluation of the residue of Rayleigh poles is comparatively simple, while the evaluation of the branch line integrals is rather complex. Therefore, Harkrider (1964) proposed « normal mode solution » in which the displacement is expressed as a sum of Rayleigh poles only, neglecting the branch line integrals [17]. For a vertically oscillating source on the surface of elastic, homogeneous and isotropic half-space, Miller and Pursey (1955) showed that two thirds of the total energy are converted in Rayleigh waves, while the remaining part goes to body waves [17].

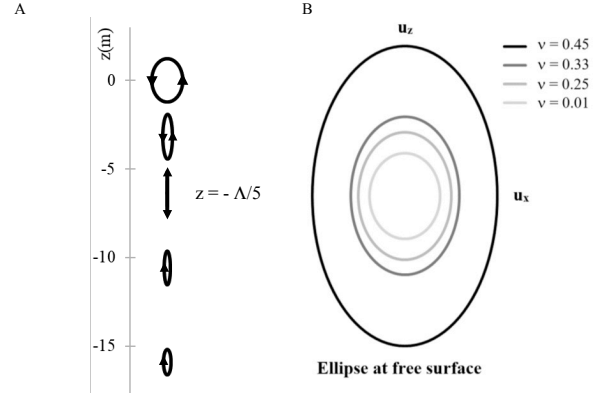

FIG. 1. (A) Example of a diagram of particle's motion on the vertical plane for Rayleigh waves at different depths. Here $\upsilon = 0.25$, $f = 10\,Hz$, $\Lambda_R = 31.8\,m$. (B) Parametric dimensionless study of ellipse pattern versus Poisson's ratio value.

Otherwise, the surface waves are proportionally attenuated with the square root of the distance versus the square of the distance for the body waves along the surface (Ewing et al., 1957) [18]. Thus, Rayleigh waves become an order of magnitude greater than that of body waves if the distance from the source exceeds three times the wavelength [20]. Experimentally we observed this phenomenon during the 2012 full-scale test [1]. In Figure 4 of the Supplemental Material, we have schematically represented the different types of waves produced by a punctual and vertical impact at the free surface of the Earth's surface. This is typically the case of dynamic compaction, which is a ground densification technique [11], with the fall of a mass of several tons. The radiation diagram for pressure waves shows a slight inclination with respect to the horizontal plane $(x, y)$. Thus we assume that the horizontal component of the motion is very strong at a distance close to the impact and in our case, it can be several tens of meters.

The full Green's function $g(r, z/z_0, \omega)$ describing the displacement field due to Rayleigh and bulk waves radiated by a force source located at $(r = 0, z = z_0)$ and satisfying the momentum equation is developed in §2 of Supplemental Material.

## IV. Description of field test

The experimental grid of the test held in 2012 near Lyon in France, is made of 23 holes distributed along five discontinuous lines of self-stable boreholes 2 m in diameter (FIG. 2). The depth of the boreholes is 5 m and the grid spacing is 7 m. The velocity of P wave was estimated between 600 et 650 m/s for the earth material near the surface. The artificial source consists of the fall of a 17 tons steel pounder from a height of about 12 m to generate clear transient vibrations pulses. In this case the depth of the source is about $z_0 = 3$ m (see Supplemental Equations), in a crater made after 6 successive impacts. The typical waveform of the source in time-domain looks like a second order Ricker wavelet. In this letter, we present data recorded for the source 1 located 30 m from the long axis of the grid. As suggested by Aki and Richards [15], this is a "near seismic field" because all the sensors are located within three wavelengths of the source and that is decisive for the interpretation of the experimental results. The signal is characterized by a mean frequency value at 8.15 Hz. The grid is mainly "sub-wavelength". At 5 m from the impact, the peak ground acceleration is around $0.9\,g$ (where $g = 9.81$ m. s$^{-2}$ is the gravity of Earth), which is significant but necessary to compensate for the strong attenuation versus distance in soils (see §3 of the Supplemental Material for more details on sensors). On each seismogram we identify, in time domain, the signature of Rayleigh waves in the global duration of the main signal (0.375 s [1]). After a few iterations on orbit graphics, this corresponds to several unambiguous criteria: a transient signal which is more energetic, a duration of 0.1 to 0.15 s, an ellipsoidal motion identified in at least two of the three planes of space, the predominance of the motion in the $(x, z)$ plane.

In frequency domain, we basically considered the grid of holes as a filter without any consideration of initial soil properties (wave velocity, pattern of the grid, etc.). We have simply calculated the magnitude in dB of three transfer functions $|T_x(\omega)|$, $|T_y(\omega)|$ and $|T_z(\omega)|$ as the spectral ratio of the ground particle velocity for a couple of sensors (A, B) (§3 and Figure 5).

$$\left|T_{B/A}(\omega)\right| = \left|\frac{\mathcal{B}(\omega)}{\mathcal{A}(\omega)}\right| \qquad (1)$$

With $\mathcal{A}(\omega)$ and $\mathcal{B}(\omega)$ the Fourier transforms of signal recorded with sensor A and B. As shown in FIG. 4, we calculated $T_{B/A}$ and $T_{C/B}$, the seismic signal passing through the grid from the right to the left of the array of holes.

## V. Experimental results with an emphasis on elliptical polarization

In time domain we present the results of the preliminary land streamer test (yellow line in Figure 5 of Supplemental Materials) and we have selected three seismograms for an extraction of the Rayleigh waves: B1, outside the grid of holes, C1 at the edge of the grid and F1 inside the grid (FIG. 2). In Figure 6 (Supplemental Materials) we present the different components of the particle's velocity recorded. For a given sensor and for the same time sampling, data are normalized by the maximum of $v_x(t)$, $v_y(t)$ or $v_z(t)$. Results presented are the projection of the particle's velocity on three planes: $(x, z)$, $(x, y)$ and $(y, z)$. Supplementary Figure 6(A) shows the ellipses of the landstreamer sensors located at 10, 20, 50 and 70 m from the source and Supplementary Figure 6(B) presents those of sensors B1, C1 and F1. At the impact point, the signal emitted is less than 0.4 s and we have identified for all ellipses, 1 to 2 cycles included in a duration of 0.1 to 0.15 s.

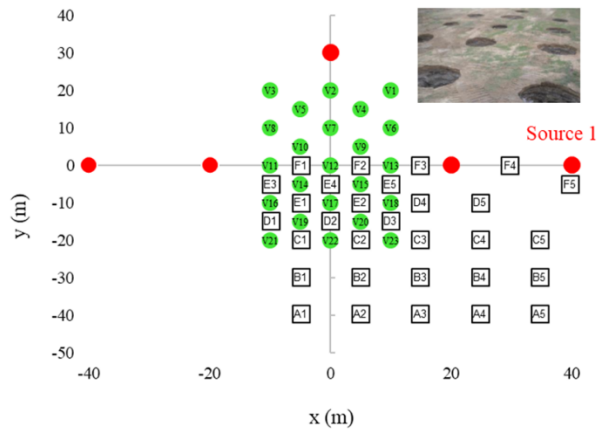

FIG. 2. Plan view of field-site layout [1]. Green disks are cylindrical holes, red disks are different source locations (S1 is tested here) and squares are the seismic sensors. Top right, the mesh of holes.

We select a set of two doublets: a sensor located between the source and the grid compared to a sensor inside the grid and this latter compared to a sensor located behind the mesh [22]. Thanks to a preliminary land streamer test with seismic sensors located in a single line (coordinates $(x = 20, y = 0)$, $(-10, 0)$ and $(-30, 0)$), before digging holes we can compare the transfer functions before and after the structuration of the soil (See Supplemental Materials, Figure 5). In Supplementary Figure 6(A) we can observe for sensors at 10 and 20 m that the orbits are subhorizontal i.e. vibrating mainly in the plane $(x, y)$ with an elongation according to $x$. The orbits are gradually verticalized with the distance (50 and 70 m). Compared with the theoretical case of vertical ellipses for Rayleigh waves, this suggests that this characteristic regime of particle's motion at the surface, only takes place after a distance of one wavelength for P-waves. From Supplementary Figure 6, we obtain identifiable ellipses of contrasted geometries and one should notice that the directions of the main axis $(a, b, c)$ conventionally described in the bottom part of FIG.3, and the ratios $a/c$ and $a/b$ can drastically change. Thus we adopt a plane view to represent the orientation and the geometry of the intersection of the ellipsoid with the horizontal $(x, y)$ plane (FIG. 3 - top). In FIG. 3, the ellipses which are the representation of the intersection of ellipsoids with the horizontal plane, provide us with three main pieces of information in terms of vertical orientation, particle's velocity magnitude and ellipse's azimuth in the horizontal plane. A fourth piece of information is the prograde or retrograde motion (Figure 6 of Supplemental Material), but we observe that is quite unstable. We decided not develop this information. At first sight, there is a great heteroclicity of ellipses, both in geometry but also in amplitude. We recall that we expect ellipses with their major axis oriented vertically but here we observe a 90 ° rotation of this axis for the majority of ellipses located between the grid and the source but also in the grid. That is the essential of the vibration is horizontal at the free surface and the dark red coloration of the ellipses illustrates this information, according the graphic convention of FIG. 3. Behind the mesh of holes and below, i.e. in the left and the bottom parts of the FIG. 3, the geometry of the ellipses in the vertical plane is either close to that of a circle (black color) or showing the verticalization of their main axis (blue color). In terms of amplitudes, the ellipses are much more energetic inside the grid and between the grid and the source, than elsewhere. We know that real soils strongly attenuate the signal energy with the offset [1] but here we notice a concentration of energy inside the grid. Two "shadow zones" are identified: at the top left of the figure and in the bottom right quarter. We are unable to identify azimuth families as clearly as for the other parameters. However, the overall motion of the ellipses suggests a double rotation of the horizontal axes of the ellipses in $(x, y)$ plane: at both y-interfaces of the grid of holes. Additional information can be obtained from the preliminary seismic test because ellipses for sensors at 10 and 20 m are mainly oriented in $x$-direction and the y-component is very tiny. The test with holes causes tilting of ellipses axes and involves a second horizontal component in the area before the grid.

Spectral analysis (FIG. 3) brings the following elements. We present the results for three sensors (see § IV). The transfer function $T_{B/A}$ represents information relating to the signal entering in the grid of holes from the right (FIG. 4) and $T_{C/B}$, that of the signal going out of the device by the left.

We have calculated the magnitude in dB of the transfer function (Eq. 1) for the initial soil (solid blue line) and for the structured soil with holes (solid red line). We have also drawn the magnitude of the original soil transfer function minus 3 dB (gray dotted line) to illustrate the efficiency of the holey-ground. We consider that results acquired with the land streamer (solid blue line) is the reference curve to compare with the others. It is the spectral signature of the soil with its initial peculiarities. The other aspects of the energy distribution in the structure (effective negative index and flat lensing, seismic metamaterials, dynamic anisotropy) have been already discussed [1].

For $T_{B/A}$ in left column of FIG. 4, results show a significant change of the shape of the magnitude in range of frequency 1-9 Hz for the 3 components of the sensor. In the horizontal $(x, y)$ plane, amplification is overall observed for 1 to 4 Hz, and de-amplification for 4 to 7 Hz. For z-component, the amplification is observed for 2 to 7.5 Hz and the de-amplification for 1 to 2 Hz and 8 to 9 Hz. For $T_{C/B}$ in right column of FIG. 9, one can notice a de-amplification for the broadband between 1 to 5 Hz and between 1 to 9 Hz respectively for the horizontal components x and y of the sensors. For $z$-component, the de-amplification is observed for 0.5 to 3.2 Hz. Roughly speaking, the signal is much more attenuated horizontally and for a broader range of frequency at the output than at the input of the mesh of holes.

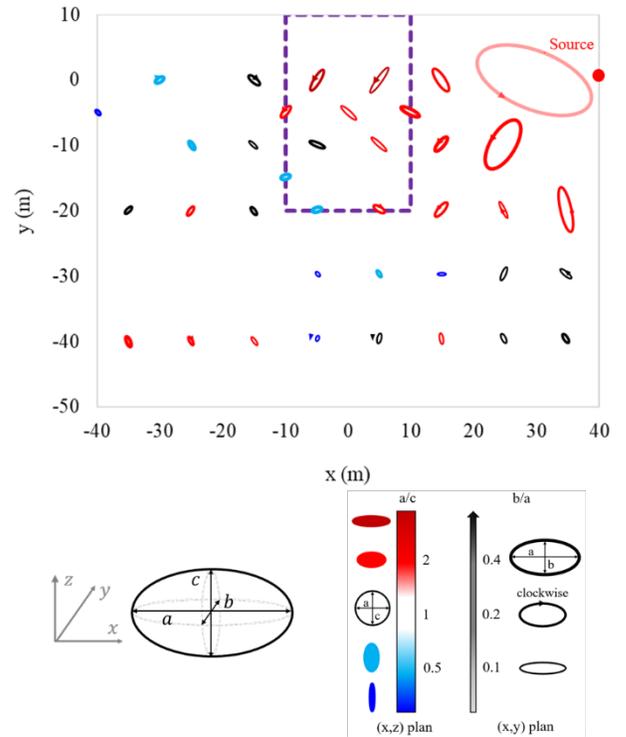

FIG. 3. Top view of ellipses with graphic conventions for ellipsoid and ellipses at the bottom part of the figure. Source is located at $(x = -40, y = 0)$. The dotted-line purple rectangle represents the part of the grid of holes

coinciding with sensors location. In this representation, we can observe the azimuth of the major axis of the ellipsoid intersecting the horizontal plane. Bottom left, the three axes of the ellipsoid: $a$ and $c$ are respectively the largest axis in $(x,y)$ and the smallest one in the $(x,z)$ plane; $b$ is the intermediate one in $(x,y)$. Bottom center, the pattern of the ellipses in the $(x,z)$ plane ($z=0$) is shown as $(a/c)$. In blue, the ellipses are stretched upwards and in dark red, in the horizontal plane. Bottom right, the thickness of the line also indicates their proportions $b/a$ in the horizontal plane.

## VI. Discussion on transfer function and velocity contrast

On the basis of data acquired on the shape of the ellipsoidal movement of the surface waves and on the spectral study showing the amplifications of the $x$, $y$ and $z$ components according to the frequency, we suggest correlations and interpretations. Thanks to geological features, we make the assumption of a half-space as soil model. We have identified the signature of the surface waves on the seismograms resulting from a vertical impact source. Theoretical ellipses can be compared to recorded data. The shape and distribution of the ellipses show that the grid of holes does not only modify the energy distribution, which is already knew from previous studies [1], but also changes the polarization of the seismic waves.

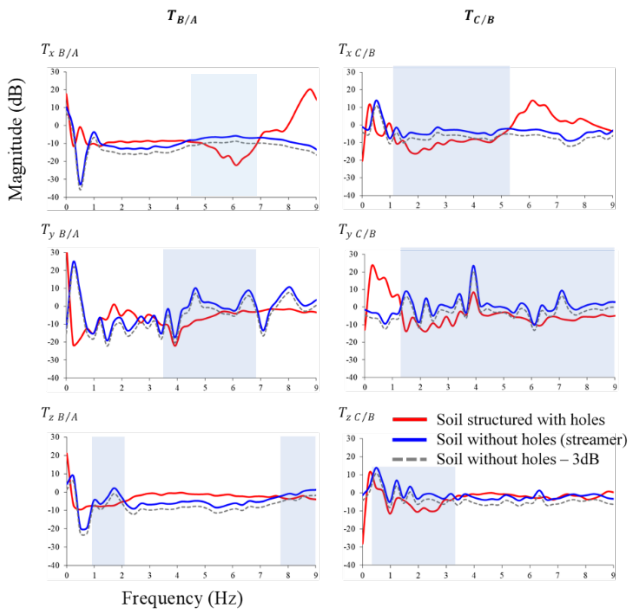

FIG. 4. Transfer functions $T_{B/A}$ and $T_{C/B}$ for $x$, $y$ and $z$ component of the sensors [22]. The light blue bands on the graphs show the frequency ranges for which a de-amplification is measurable after digging holes.

Indeed we observe a clear amplification of the horizontal-component of particle's velocity inside the grid with a tilting of most of the ellipses. Theoretically vertical, the major axis has turned $\pi/2$ for the preliminary test but also for the ground with holes. The land streamer shows that below a wavelength for the mean frequency, we are still in "near-field" (vicinity of the source) with an unstabilized particle's motion regime. For the ground with holes, this could be correlated with the horizontal amplification shown by $T_{x_{B/A}}$ and $T_{y_{B/A}}$ functions for frequency $f < 4Hz$. We give two explanations to this observation. First, the right border of the grid of holes is acting like a seismic reflector, by analogy with optics, and changes the wave polarization. This interpretation could also explain the thickening ($b/a$) of the ellipses in $(x,y)$ plane, located between the source and the grid, as the combination of two wave fields, an incident and a reflected one.

Various authors ([23, 24, 25]) describe the influence of a horizontal layer over half-space (thickness $H_1$) and the velocity contrast ($V_2/V_1$) on the ellipticity of Rayleigh waves. There is a frequency dependence of wave motion (prograde or retrograde) and then on the ellipse's shape. In our case, we could imagine such a situation with the existence of two layers (initial soil and structured soil) but with a vertical interface. For the second explanation, we propose that the local resonance of the holes causes an energy conversion by horizontal vibration modes of the holes in the $(x,y)$ plane. The vertical amplification is non-null inside the mesh. We suggest that the cause is the vertical vibration mode of the holes. Another way to interpret this data in the grid and in the area between the grid and the source, is to invoke a Love mode conversion. In the area located at the left of the grid of holes, the major axis of the ellipses is mainly verticalized. This can be interpreted as the crossing of a second reflector (the left border of the mesh) and then a change of polarization or the consequence of a strong energy dissipation in the $(x,y)$ plane inside the grid itself. However, regardless of the interpretation of the physical phenomena, there is a clear attenuation of the horizontal component of the motion at the exit of the grid. We believe this is an essential result in the idea of protection against vibrations.

## VII. Perspective on seismic metamaterials with dynamic effective elasticity tensors

The full-scale experiment on a subwavelength device near the city of Lyon has already provided valuable information on the distribution of seismic energy [1]. In the present article, we show the complexity of the wave propagation in a structured soil by analyzing the surface particle's velocity in 3D. The recorded data suggest various physical phenomena take place, including changes in polarization or wave mode conversions. For the civil engineering purpose, a strong attenuation of the horizontal component of the motion is observed behind the grid at the opposite side of the seismic source. Otherwise, we can suspect the holes in the ground act as Helmholtz resonators with large aperture [26] especially for a vertical stress. According to the geometry of real cavities in this experiment, the theoretical Helmholtz frequency of a single hole, is of the order of 10 Hz. This article also raises the concept of media with effective properties. Indeed, we could interpret the change of ellipse shape as the consequence of an effective Poisson's ratio (FIG. 5 and FIG.3 and more detail in §5 of Supplemental Material) and this opens interesting avenues in the design of seismic metamaterials with tunable Poisson's ratio similarly to what has been experimentally achieved with pentamode metamaterials at the micrometer scale [13], with potential applications far beyond those of electromagnetic metamaterials [27]. Indeed, large scale mechanical metamaterials with auxetic properties could find applications in earthquake protection [28]. Finally, we point out that our experimental data can be interpreted in terms of cloaking of Rayleigh waves via anomalous resonances [29] induced by holes. Indeed, considering as in [1] that the soil periodically structured with boreholes behave like an effective medium with strong dynamic anisotropic features (such as hyperbolic media), one can invoke space folding arguments to reinterpret data in Fig.4 as being some form of cloaking for the source. This concept which is still in its infancy and thus beyond the scope of the present article, would find applications in seismology similar to those proposed in [30] in electromagnetics.

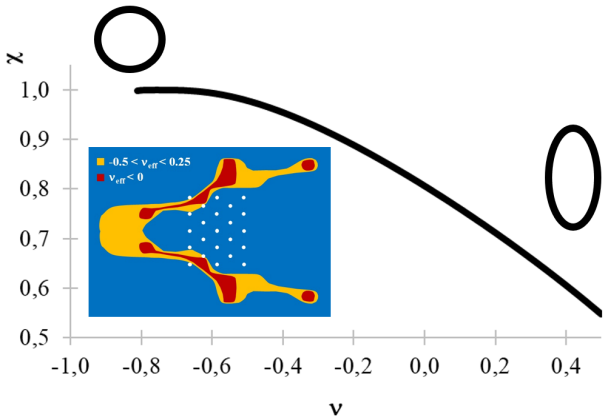

FIG. 5. Horizontal to vertical displacement ratio χ at the free surface versus Poisson's ratio. Cartography with outlines drawn around the sensors for which the geometry of the ellipses can be interpreted as a negative Poisson's coefficient.

It is interesting to note that conversion of Rayleigh waves recently studied with small scale experiments [31, 32] could well be revisited to unveil radical changes of Rayleigh wave ellipticity in forests of trees [33], in light of our findings. We believe the future of seismic metamaterials is bright, and we hope the extreme dynamic effective features reported here for soils structured with holes, will lead to further analogies with composite structures such as those introduced by Milton and Cherkaev more than twenty years ago [34]. Structured soils could then potentially match any prescribed elasticity tensor in a dynamic regime, and this would be a natural path to a seismic cloak.

## Acknowledgements

SG is thankful for a visiting position in the Mathematics Department at Imperial College London supported by EPSRC program grant EP/L024926/1.